# Governing frontier general-purpose AI in the public sector: adaptive risk management and policy capacity under uncertainty through 2030

**Fabio Correa Xavier**

fabio@fabioxavier.com.br

www.fabioxavier.com.br

www.linkedin.com/in/fabiocorreaxavier

## Abstract

The governance of frontier general-purpose artificial intelligence has become a public-sector problem of institutional design, not merely a technical issue of model performance. Recent evidence indicates that AI capabilities are advancing rapidly, though unevenly, while knowledge about harms, safeguards, and effective interventions remains partial and lagged. This combination creates a difficult policy condition: governments must decide under uncertainty, across multiple plausible trajectories of progress through 2030, and in environments where adoption outcomes depend on organizational routines, data arrangements, accountability structures, and public values. This article argues that public governance for frontier AI should be based on adaptive risk management, scenario-aware regulation, and sociotechnical transformation rather than static compliance models. Drawing on the *International AI Safety Report 2026*, OECD foresight and policy documents, and recent scholarship in digital government, the article first reconstructs the conceptual foundations of the "evidence dilemma," differentiated AI risk categories, and the limits of prediction. It then examines how AI adoption in government depends on organizational redesign, public-sector institutional dynamics, and data collaboration capacity. On that basis, it proposes an adaptive governance framework for public institutions that integrates capability monitoring, risk tiering, conditional controls, institutional learning, and standards-based interoperability. The article concludes that effective AI governance requires stronger policy capacity, clearer allocation of responsibility, and governance mechanisms that remain robust across divergent technological futures. [1][2][3][4]

## 1  Introduction

Frontier general-purpose AI has evolved from narrow conversational assistance to systems capable of high performance in coding, scientific reasoning, image generation, and other complex tasks. Yet the policy significance of this shift lies not only in higher benchmark scores. It lies in the fact that governments, regulators, and public organizations must now govern technologies whose capabilities evolve faster than the evidence base regarding their risks, limits, and societal effects. The *International AI Safety Report 2026* names this mismatch the "evidence dilemma": acting too early may entrench weak or miscalibrated interventions, while waiting for more complete evidence may leave societies exposed to material harms. [1]

This problem becomes more acute because there is no agreed single trajectory for AI progress through 2030. OECD scenario work explicitly treats future AI development as a set of plausible but uncertain futures rather than a forecast with assignable probabilities. Its scenarios range from plateau and slowdown to continued or accelerated progress, and the OECD notes that current evidence is insufficient to rule out any of them. [2]

For public administration, the relevance of this problem is immediate. Public institutions are under pressure to adopt AI for efficiency, service delivery, inspection, adjudication support, fraud detection, and policy analysis. At the same time, recent digital government research shows that public AI adoption is not a simple matter of procurement or software deployment. It requires changes in routines, structures, governance arrangements, data practices, and organizational culture. [5][6][7][8]

The central thesis of this article is that governance for frontier general-purpose AI in the public sector should be built around adaptive risk management under uncertainty. Such governance must combine three elements: differentiated treatment of distinct risk classes, sociotechnical transformation within public organizations, and policy and regulatory architectures that remain robust across multiple plausible AI futures. The remainder of the article is organized as follows. Section 2 presents the theoretical foundation. Section 3 develops the analytical discussion. Section 4 proposes a practical implementation framework. Section 5 examines implications for leadership, governance, regulation, and public policy. Section 6 connects the argument to broader standards and reference models. Section 7 closes with limitations and a future agenda. [1][2][5][6]

## 2. Theoretical foundation

### 2.1. General-purpose AI as a cross-sector governance object

General-purpose AI should be understood as an infrastructural and cross-context technology rather than as a single-purpose application. Its governance is therefore structurally different from that of sector-specific digital tools. The *International AI Safety Report 2026* emphasizes that general-purpose systems can be deployed across multiple sectors and workflows, which complicates evaluation, oversight, and assignment of responsibility. [1]

A further complication is that capability growth can no longer be explained solely by pre-training scale. The report notes substantial recent gains from post-training methods and inference-time scaling, especially in mathematics, software engineering, and scientific reasoning. This means that regulatory approaches tied only to one upstream variable, such as training scale, are likely to be incomplete. [1][2]

### 2.2. The evidence dilemma and adaptive governance

The evidence dilemma is a central policy problem. In the public sector, it implies that governance cannot depend on waiting for definitive proof about every risk before acting. At the same time, it cannot justify broad, undifferentiated restrictions based on speculative claims alone. A defensible theoretical response is adaptive governance: institutions should act on the best available evidence, explicitly state uncertainties, monitor outcomes, and revise controls as evidence accumulates. [1][2]

This position is reinforced by the OECD's strategic foresight method. The trajectories paper uses trend analysis, horizon scanning, driver mapping, and technology road mapping to build scenarios that are meant to inform policy rather than predict the future. That methodological choice matters because it shifts governance away from deterministic planning and toward robustness-oriented planning. [2]

### 2.3. Differentiated AI risk categories

The *International AI Safety Report 2026* groups general-purpose AI risks into malicious use, malfunctions, and systemic risks. This typology is analytically important because it prevents a conceptual collapse of very different harm mechanisms into one generic category. Malicious-use risks include scams, fraud, cyber abuse, and harmful content generation. Malfunction risks concern unreliability, brittle behavior, and unsafe outputs in operational settings. Systemic risks include labor disruption, concentration of power, erosion of autonomy, and cumulative institutional dependence. [1]

A differentiated risk typology implies differentiated governance. Some risks justify immediate operational safeguards because harms are already documented. Others are still emerging and therefore require threshold-based monitoring, scenario triggers, and precaution without false precision. [1][4]

### 2.4. Digital government as a sociotechnical domain

The digital government literature is especially relevant here because it cautions against narrow technological determinism. A long-standing review in *Government Information Quarterly* argued that e-government research often suffered from definitional vagueness and oversimplified development models, calling for more grounded, context-sensitive explanation of public-sector digitalization processes. [9]

Recent work extends that critique to AI. Tangi, Rodriguez Müller, and Janssen argue that AI integration in public administration is a sociotechnical phenomenon in which benefits depend on changes in routines, organizational structures, governance, and culture, not merely on technical installation. Busch, Johannessen, and Pekkola similarly show that digital government remains theoretically fragmented and overly dependent on borrowed theories, which limits cumulative understanding of public-sector digital phenomena such as accountability, public value, trust, and explainability. [5][6]

## 3. Analytical development

### 3.1. Rapid but uneven capability progress

Current progress in frontier AI is significant, but uneven rather than smooth. The safety report states that capabilities are improving rapidly yet remain "jagged," with leading systems excelling in difficult domains while still failing on tasks that appear simpler, such as some forms of counting, spatial reasoning, or longer error recovery. This has direct implications for public decision-making. Benchmark excellence cannot be treated as equivalent to institutional reliability. [1]

The OECD trajectories paper adds an operational lens by examining how long AI systems can sustain task completion at specified success rates across domains such as scientific reasoning, software engineering, web navigation, and robotics. Even though the paper states that the resulting projections are illustrative, not predictive, the task-horizon perspective is important for governance because it shifts attention from isolated performance to sustained autonomous action. [2]

### 3.2. Materializing risks and evidentiary asymmetries

The safety report indicates that several harms are already materializing, including scams, fraud, blackmail, non-consensual imagery, and some cyber-related misuse. Yet it also stresses that systematic data on prevalence and severity remain limited. This asymmetry between visible incidents and incomplete measurement is typical of fast-moving technological environments. Public institutions therefore face a double challenge: they must neither dismiss harms because the datasets are incomplete nor overstate empirical certainty where it does not exist. [1]

The OECD paper on media-reported AI incidents reinforces this point. It identifies increasing coverage in areas such as synthetic media, child safety, cyberattacks and fraud, and labor-market disruption, while also showing that media salience varies substantially across themes and over time. Media reporting is not a complete measure of objective harm, but it can serve as an early-warning signal for issues shaping public attention and policy agendas. [4]

### 3.3. Public-sector AI adoption is organizational, not merely technical

A critical analytical mistake is to assume that public-sector AI outcomes depend mainly on model quality. Recent digital government research suggests otherwise. Tangi et al. argue that government AI adoption requires transformation of routines and practices into new sociotechnical forms, and that benefits are realized only when changes occur in structure, governance, and culture. [5]

This conclusion is consistent with research on organizing public-sector AI adoption, which finds that public organizations manage tensions between innovation and established institutional identity through different structural arrangements, including separated data teams and more integrated approaches, each associated with distinct barriers and support processes. In practical terms, AI adoption fails not only because of technical limitations, but also because organizations struggle to reconcile innovation with public accountability, legal obligations, and established administrative routines. [8]

## 3.4. Data collaboration and governance capacity

The expansion of AI in government increases the importance of data-sharing and data-governance arrangements. Schmeling, al Dakruni, and Mergel show that data collaboration in digital government is multidimensional, involving ecosystem, organizational, and individual levels. Their review also finds a decline in attention to standardization and data management relative to innovation and participation, despite the importance of those elements for federated data ecosystems. [7]

This finding is highly relevant to frontier AI governance. Public organizations cannot implement reliable, auditable, and explainable AI without robust data stewardship, interoperable metadata, clear sharing rules, and institutional mechanisms for coordination. In that sense, AI governance and data governance are not separate agendas. The former depends materially on the maturity of the latter. [3][7]

## 3.5. Why static regulation is insufficient

The safety report notes that frontier AI risk management has become more structured in 2025 and 2026 through company safety frameworks, transparency initiatives, code-of-practice instruments, and incident reporting arrangements. At the same time, it emphasizes that evidence on real-world effectiveness remains limited, standardization is incomplete, and voluntary frameworks vary in scope, thresholds, and enforceability. [1]

This reveals a core limitation of static regulation. When capabilities change through multiple technical pathways and deployment arrangements, a one-time rule set can quickly become mismatched with practice. Regulatory design should therefore privilege transparency, evaluation, incident reporting, auditability, threshold-based escalation, and periodic revision rather than purely static obligations. [1][2]

# 4. Practical section: an adaptive governance framework for public institutions

This article proposes the Adaptive Public-Sector Frontier AI Governance Framework, intended for ministries, courts, oversight bodies, regulators, public enterprises, and large administrative agencies.

## 4.1. Foundational premise

The framework assumes that institutions cannot rely on any single forecast of AI progress. Governance should therefore be designed to remain usable under at least three conditions: progress slows, progress continues at roughly current rates, or progress accelerates substantially. [2]

## 4.2. Six operational layers

### 4.2.1. Capability intelligence

Public institutions should maintain a structured capability-monitoring function covering autonomy horizon, tool use, multimodal performance, inference-time reasoning, and domain-specific reliability. The purpose is not technology monitoring for its own sake. It is to detect when new forms of harm, dependency, or delegation become plausible. [1][2]

### 4.2.2. Risk tiering

Each use case should be classified by risk type, sector sensitivity, and reversibility of error. At minimum, institutions should distinguish misuse risk, malfunction risk, and systemic risk, then overlay sectoral criticality such as justice, health, taxation, procurement, inspection, benefits administration, and cybersecurity. [1]

#### 4.2.3. Conditional controls

Where uncertainty is high, governance should rely on predefined "if-then" triggers. If certain capabilities, incident patterns, or misuse indicators emerge, then stricter access controls, independent review, enhanced human oversight, or suspension rules should activate automatically. This follows the report's discussion of conditional safeguards in frontier safety frameworks. [1]

#### 4.2.4. Defense-in-depth

No single safeguard should be treated as sufficient. A layered architecture should combine model controls, role-based access, workflow review, logging, provenance measures, red-teaming, incident reporting, and post-deployment monitoring. The safety report explicitly identifies defense-in-depth as a means of improving robustness where individual safeguards remain limited. [1]

#### 4.2.5. Sociotechnical implementation

Every high-impact AI project should include an organizational implementation and redesign plan. This plan should specify workflow changes, accountability assignment, training requirements, review rights, override mechanisms, data responsibilities, and documentation duties. The public value of AI depends on these institutional arrangements, not only on technical performance. [5][8]

#### 4.2.6. Learning and revision cycle

Governance arrangements should be reviewed on a fixed cycle, such as quarterly or semiannually, with mandatory review of incidents, near misses, model updates, procurement dependencies, and changes in task autonomy. This is the practical implication of governing under uncertainty: institutions must learn continuously, not merely certify once. [1][4]

Figure 1 presents the six operational layers of the proposed framework and illustrates their cyclical relationship in adaptive public-sector AI governance.

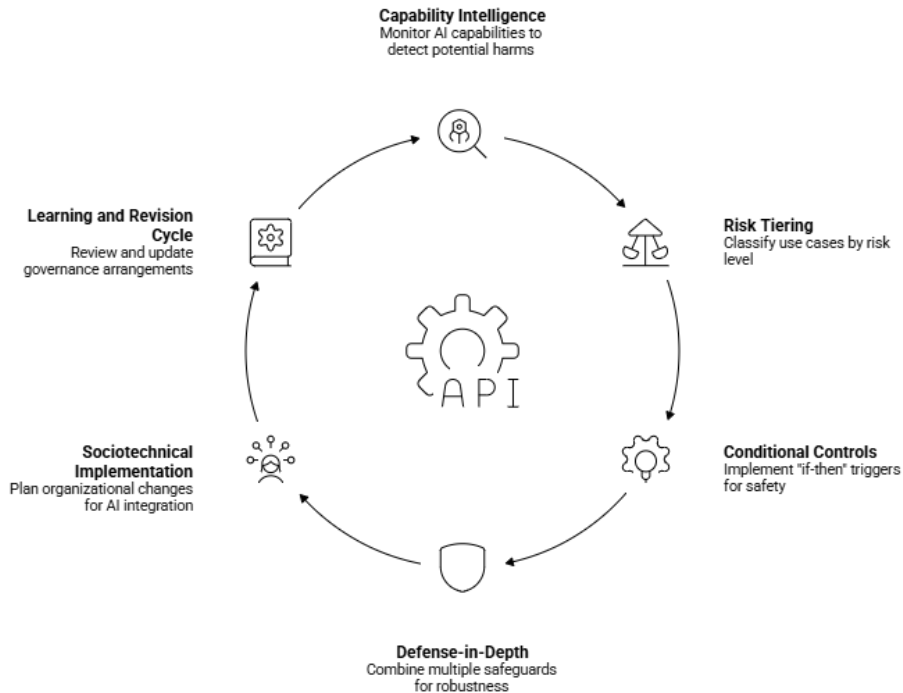

*Figure 1 - Six operational layers*

# 5. Implications for leadership, governance, regulation, and public policy

### 5.1. Leadership

Leadership in AI governance requires institutional judgment rather than generalized enthusiasm for innovation. Senior officials must define risk appetite, approve escalation thresholds, allocate resources for monitoring, and protect meaningful opportunities for human review and contestation in consequential decisions. Where these responsibilities remain diffuse, AI adoption tends to oscillate between underuse and uncontrolled experimentation. [1][5]

### 5.2. Governance and accountability

The safety report repeatedly highlights information asymmetries, fragmented visibility across the AI value chain, and uncertainty about which actors are best positioned to mitigate which risks. This is especially problematic in government, where public responsibility cannot be displaced onto vendors or technical subcontractors. Governance frameworks should therefore map roles explicitly across developers, integrators, deployers, policy owners, security functions, legal oversight bodies, and audit functions. [1]

### 5.3. Regulation

Regulation should be adaptive, auditable, and proportionate. The emerging landscape described in the safety report, including transparency obligations, incident reporting mechanisms, and general-purpose AI governance initiatives, suggests that effective regulation will depend less on one exhaustive statute and more on interoperable governance instruments. These include documentation, evaluation standards, reporting duties, and review processes linked to risk and capability thresholds. [1]

### 5.4. Public policy capacity

The OECD.AI Index is particularly relevant here because it conceptualizes national implementation of trustworthy AI through both policy actions and observable ecosystem outputs. It organizes assessment around dimensions such as research and development, infrastructure, governance, accessibility, and policy environment. Although the index is still evolving, its design implies that AI readiness is not reducible to one legal instrument or one technology program. It is, instead, a composite capacity problem. [3]

# 6. Integration with broader norms, frameworks, and reference models

The OECD materials reviewed here, together with recent digital government literature, point toward a broad but coherent normative architecture. First, trustworthy AI governance requires policy interoperability rather than isolated national improvisation. The OECD.AI Index is explicitly anchored in the OECD Recommendation on Artificial Intelligence and measures the implementation of national policy recommendations for trustworthy AI. [3]

Second, public-sector AI governance must be linked to public-administration values that generic private-sector adoption models often understate. Recent work in digital government theory argues that the field needs stronger concepts tailored to public values, accountability, trust, explainability, sustainability, and ethical concerns specific to government settings. [6][9]

Third, the design of AI governance should be compatible with composite and comparative assessment tools. The OECD.AI Index notes that cross-country policy measurement is difficult because of limited comparable data and rapidly changing AI-related inputs, outputs, and policy interventions. This caveat is important: standards and indicators are necessary, but they should not be mistaken for complete representations of real-world governance quality. [3]

## 7. Final considerations

This article has argued that frontier general-purpose AI governance in the public sector should be approached as a problem of adaptive institutional design under uncertainty. The evidence reviewed here supports four main conclusions. First, frontier AI capabilities are advancing quickly, but unevenly. Second, public risks are heterogeneous and should be differentiated analytically and operationally. Third, AI adoption in government is sociotechnical and depends on organizational redesign, data collaboration, and accountability structures. Fourth, robust public governance cannot depend on one forecast of technological progress through 2030. [1][2][5][7]

The proposed framework responds to these conditions by combining capability intelligence, differentiated risk tiering, conditional controls, defense-in-depth, sociotechnical implementation, and iterative revision. Its practical value lies in helping public institutions move beyond two inadequate positions: passive waiting for certainty and uncritical acceleration driven by procurement or political pressure. [1][4]

This article has important limitations that should be stated clearly. It is a synthesis and framework-building exercise grounded in recent institutional reports and selected peer-reviewed literature. It does not provide original empirical testing of the proposed model, nor does it settle contested questions about long-term AI trajectories, the comparative effectiveness of safeguards, or labor-market effects. Those remain active research agendas. [1][2][6]

A plausible next agenda for research and practice includes stronger public incident-reporting systems, more realistic deployment evaluations, comparative studies of organizational arrangements for public AI adoption, and closer integration between AI governance and data-governance infrastructures. In institutional terms, the key question is no longer whether uncertainty exists. It is whether public institutions can govern responsibly while uncertainty persists. [1][2][7][8]